\documentclass{article}
\usepackage{graphicx}
\usepackage{dcolumn}
\usepackage{bm}

\usepackage[utf8]{inputenc}
\usepackage[english]{babel}
\usepackage[T1]{fontenc}
\usepackage{amsmath}
\usepackage{hyperref}

\usepackage{tikz}
\usepackage{lipsum}

\usepackage[linguistics]{forest}
\usepackage{tikz-cd}
\usepackage{subcaption}

\usepackage[a4paper,top=3cm,bottom=2cm,left=3cm,right=3cm,marginparwidth=1.75cm]{geometry}

 \usepackage{bm}

 \usepackage{graphicx}
 \usepackage{url}
 \usepackage{braket}

 \usepackage{listings}
 \usepackage{mathtools}

 \usepackage{amsthm}
 \usepackage{amsfonts}
 \usepackage{amssymb}
 \usepackage{amsmath}
 \usepackage{qcircuit}
 \usepackage{enumerate}

 \usepackage{algpseudocode}
 \usepackage{algorithm}
\usepackage[style=base]{caption}
\usepackage{authblk}

\newtheorem{theorem}{Theorem}
\newtheorem{corollary}{Corollary}

\DeclareMathOperator*{\argmin}{arg\,min}

\newcommand{\norm}[1]{\left\lVert#1\right\rVert}

\title{Classification of MNIST dataset with Quantum Slow Feature Analysis}

\author[1,3]{Iordanis Kerenidis}
 
\author[2,3]{Alessandro Luongo}%

\affil[1]{QC Ware Corp. -  Palo Alto, USA}
\affil[2]{Atos Quantum Lab - Les Clayes sous Bois, France}
\affil[3]{CNRS, IRIF, Université Paris Diderot - Paris, France}

\date{\today}

\begin{document}

\maketitle
\begin{abstract}
Quantum machine learning is a research discipline intersecting quantum algorithms and machine learning. While a number of quantum algorithms with potential speedups have been proposed, it is quite difficult to provide evidence that quantum computers will be useful to solve real-world problems. Our work makes progress towards this goal. In this work, we design quantum algorithms for dimensionality reduction and for classification, and combine them to provide a quantum classifier that we test on the MNIST dataset of handwritten digits. We simulate the quantum classifier, including errors in the quantum procedures, and show that it can provide classification accuracy of $98.5\%$. The running time of the quantum classifier is only polylogarithmic in the dimension and number of data points. Furthermore, we provide evidence that the other parameters on which the running time depends scale favorably, ascertaining the efficiency of our algorithm. 
\end{abstract}

\section{Introduction}
Quantum computing has the potential to revolutionize information and communication technologies, and one of the most promising areas is quantum machine learning. Recently, many quantum algorithms with potential speedups were proposed \cite{kerenidis2016recommendation, lloyd2016quantumtopological, rebentrost2014quantumsvm, liu2018quantum, wiebe2012quantum, Lloyd2013, childs2012hamiltonian}. Nevertheless, whether quantum processing machines can be used in practice to solve efficiently and accurately real-world problems remains an open question. Answering this question consists in finding practical applications of quantum computing with a real economic and societal impact. Our work provides evidence that large-scale quantum computers with quantum access to data will be indeed useful in machine learning. We design two efficient quantum algorithms: one for dimensionality reduction and the other for classification. In addition, we simulated their combined behavior on the commonly-used MNIST dataset of handwritten digits, showing that our classifier is able to provide accurate classification, comparable to classical algorithms.\\

Our first contribution consists of a new quantum algorithm called quantum Slow Feature Analysis (QSFA): a quantum method for dimensionality reduction. Dimensionality reduction (DR) is a technique used in machine learning in order to reduce the dimension of a dataset, while maintaining the most meaningful information contained therein. DR algorithms are often used both to render the computational problem more feasible, and to counterbalance the \emph{curse of dimensionality}: the undesired property of some machine learning algorithms, whose performance deteriorates once the dimension in the feature space becomes too high. Intuitively, this is because if the data lives in an excessively high dimensional space, the informative power of the data points in the training set decreases, thus leading to a degradation in classification performance. 
\\

The second contribution is a novel algorithm for classification called quantum Frobenius Distance (QFD) classifier. It is a quantum method that classifies a new data point based on the average squared $\ell_2$-distance between the test point and each labeled cluster. Being a very simple algorithm, we believe it can target quantum hardware in the early NISQ (Noisy-Intermediate Scale Quantum) architectures. \\
 
As last contribution, the simulation results of our quantum classifier  - combining QSFA and QFD -  allows us to claim that the accuracy on the MNIST data set is around 98.5\%, which is comparable to classical machine learning algorithms, and better than many of the previous results registered in \cite{lecun1998mnist}. We also provide an estimate of the running time of our algorithm by calculating all the parameters that appear on the asymptotic running time, and depend on the input data. Our estimate provides evidence that our quantum classifier can be more efficient than the corresponding classical one, or more importantly, that one can use our quantum classifier with higher input dimension, retaining efficiency and accuracy together. \\

The goal of this work is to assess the power of a quantum computer that can access classical data efficiently, in a similar manner to classical high-performance computing machines which by default possess a fast RAM. In other words, we assume the quantum computer can use an efficient quantum procedure to create quantum states corresponding to the classical data, for example through a QRAM: a data structure that allows quantum access to classical data.
Efficient algorithms for creating such QRAM circuits using quantum oracle access have been proposed \cite{Prakash:EECS-2014-211} \cite{kerenidis2016recommendation,kerenidis2017quantumsquares}. Sometimes, in literature this data structure goes under the name of KP-Trees \cite{rebentrost2018quantum}. We can imagine the QRAM in two ways. As a quantum operator, that allows classical data to be retrieved efficiently in superposition, and as a particular format we impose on the classical data we store. In fact, the circuit for the QRAM of a dataset $X$ holds all the information needed to retrieve the matrix $X$, and is efficiently built from $X$. Of course, we do not have such quantum computers or quantum RAM right now and there is a possibility that they may never be realized. Still, there are some proof-of-concept experiments in this direction \cite{jiang2019experimental,hann2019hardware, park2019circuit, bang2019optimal}. We think of our results as motivation for actually building such quantum processing machines. We will provide more details about the QRAM model in the following sections. After some review on previous work and introduction on the techniques used, we provide short description of the classical SFA technique and of the quantum procedures for performing linear algebra.\\

\paragraph{Previous work}
 We describe previous work on quantum classification. Quantum machine learning can be roughly divided in two different approaches. The first category of algorithms uses circuits of parameterized gates to perform machine learning tasks such as classification or regression \cite{verdon2017quantum, webie2018quantumcentric, benedetti2019parameterized}. In the training phase, the parameters of the circuit are learned using classical optimization techniques, where the function to optimize is a loss function calculated on the output of the quantum circuit \cite{Schuld2017interference}. Works in this direction are \cite{otterbach2017unsupervised,farhi2018classification}, with issues outlined in \cite{mcclean2018barren} and addressed in \cite{grant2019initialization}. The second approach, uses quantum computers to speed up the linear algebraic operations performed in classical machine learning, extending the famous HHL algorithm \cite{HarrowHassidim2009HHL}. The work of Rebentrost et al. \cite{rebentrost2014quantumsvm} consists of an algorithm for a Support Vector Machines classifier. In this class of algorithms, other quantum dimensionality reduction algorithm exists: Principal Component Analysis \cite{Lloyd2013} and Linear and Nonlinear Fisher Discriminant Analysis \cite{iris2016discriminant}, which are based on Hamiltonian simulation techniques. Other possible approaches consist in using quantum annealing, like \cite{Benedetti2016assistedgraphicalmodel, kaynak1995methods,mackay2003information}.

In the last year, many results in classical machine learning were obtained by ``dequantizing'' quantum machine learning algorithms \cite{tang2018quantum, tang2018quantum2}, using techniques from randomized linear algebra and MCMC (Monte Carlo Markov Chain)  \cite{frieze2004fast}. Remarkably, these class of classical randomized algorithms works under assumptions similar to the ones in their quantum counterparts (i.e. they assume the ability to have query and weighted sample access to the dataset). Similarly to what the QRAM does for quantum algorithms, these techniques store and precompute all the partial norms for the dataset, which are later used in sampling procedures. As for the quantum case, the runtime of these classical algorithms is poly-logarithmic in the dimensions of the dataset. However, these new algorithms are unfortunately impractical on interesting datasets: the most recent proposal for solving linear system of equations has a runtime of  
$\tilde{O}(\kappa^{16}k^6\norm{A}_F^6/\epsilon^6)$, where $k$ is the rank, $\epsilon$ is the error in the solutions, $\kappa$ is the condition number of the dataset, and $\norm{A}_F$ is the Frobenius norm of the dataset. These algorithms have been implemented and benchmarked on real and synthetic datasets \cite{arrazola2019quantum}. There, the authors conclude that these techniques cannot compare in terms of runtime to other approaches in classical machine learning, and therefore these algorithm are not likely to change the set of problems which are expected to be solved efficiently by quantum computers.\\

\paragraph{Classical Slow Feature Analysis}\label{c_sfa}
SFA was originally proposed as an \emph{online, nonlinear, and unsupervised  algorithm} \cite{wiskott1999learning,Berkes2005pattern, wiskott1999learning}. It has been motivated by the  \emph{temporal slowness principle}, a hypothesis for the functional organization of the visual cortex and possibly other sensory areas of the brain \cite{scholarpedia2017SFA} and it has been introduced as a way to model some transformation invariances in natural image sequences \cite{zhang2012slow}. SFA formalizes the slowness principle as a nonlinear optimization problem \cite{blaschke2004independent, sprekeler2014extension}. A prominent advantage of SFA compared to other algorithms is that it is almost hyperparameter-free. Another advantage is that it is guaranteed to find the optimal solution within the considered function space \cite{escalante2012slow,Sprekeler2008understandingframework}.  It has been shown that solving the optimization problem upon which SFA is based is equivalent to other dimensionality reduction algorithms, like Laplacian Eigenmaps \cite{sprekeler2011relation} and Fisher Linear  Discriminant \cite{klampfl2009replacing}, thus a quantum algorithm for SFA also provides algorithms for Laplacian Eigenmaps and Fisher Linear Discriminant. Our quantum algorithm runs in time polylogarithmic in the dimension and number of points in the dataset, thus with a potential considerable speedup with respect to classical algorithms. With appropriate preprocessing, SFA can be used as a dimensionality reduction algorithm to improve speed and classification accuracy in supervised machine learning \cite{Berkes2005pattern,Gu2013supervised, zhang2012slow, sun2014deepsfaactionrecognition}. The problem is formalized as follows. The input of the algorithm consists of vectors $x(i) \in \mathbb{R}^d , i \in [n]$. Each $x(i)$ belongs to one of $K$ different classes. By definition, SFA algorithm computes the $K-1$ functions $g_j( x(i)) : \mathbb{R}^d \rightarrow \mathbb{R}, j \in [K-1]$ such that the output $ y(i) = [g_1(  x(i)), \cdots , g_{K-1}(  x(i)) ]$ is very similar for the training samples that belongs to the same class and largely different for samples of different classes. Once these functions are learned, they are used to map the training set in a low dimensional vector space of dimension $K-1$. When a new data point arrives, it is mapped to the same vector space, where classification can be done with higher accuracy. We introduce the minimization problem as it is stated for classification \cite{Berkes2005pattern}.  
For $T_k$ the set of training elements of class $k$, let $a=\sum_{k=1}^K \binom{|T_k|}{2}.$ 
\noindent
For all $j \in [K-1]$, minimize: 
\begin{equation}
    \Delta(y_j) =  \frac{1}{a} \sum_{k=1}^K \sum_{\substack{s,t \in T_k \\ s<t}} \left( g_j( x(s)) - g_j( x(t)) \right)^2 
\end{equation}
with  the following constraints $\forall v < j $: 
\begin{enumerate}
	\item $\frac{1}{n} \sum_{k=1}^{K}\sum_{i\in T_k} g_j( x(i)) = 0 $
	\item $\frac{1}{n} \sum_{k=1}^{K}\sum_{i \in T_k} g_j( x(i))^2 = 1 $
	\item $ \frac{1}{n} \sum_{k=1}^{K}\sum_{i \in T_k} g_j( x(i))g_v( x(i)) = 0$ 
\end{enumerate}
In order for the minimization problem to be feasible in practice, the $g_j$'s are restricted to be  linear functions $w_j$, such that the output signal becomes $ y(i) = [w_1^T x(i), \cdots  w_{K-1}^T x(i) ]^T$ or else $Y=XW$, where $X \in \mathbb{R}^{n \times d}$ is the matrix with rows the input samples and $W \in \mathbb{R}^{d \times (K-1)}$ the matrix that maps the input matrix $X$ into a lower dimensional output $Y \in \mathbb{R}^{n \times (K-1)}$. In case it is needed to capture nonlinear relations in the dataset, one commonly performs a nonlinear polynomial expansion on the input data during the preprocessing. Usually, a polynomial expansion of degree $2$ or $3$ is sufficient, since polynomial of higher order might overfit. The constraint on the average and variance of the signal's component can be satisfied efficiently by normalizing and scaling the input matrix $X$ before solving optimization problem. This assumption is simple to satisfy, as normalizing the input matrices of the dataset is linear in the dimension of the problem. Taking all this into account, we can restate the definition of the delta function as:
\begin{equation}\label{delta}
\Delta(y_j) = \frac{ w^T_j A  w_j}{  w_j^T B  w_j} ,
\end{equation}
where the matrix $B$ is called the covariance matrix of the dataset, and is defined as:
\begin{equation} \label{defB}
B:= \frac{1}{n}\sum_{i \in [n]} x(i) x(i)^T = X^T X 
\end{equation}
and the matrix $A$ is called the derivative covariance matrix and is defined as: 
\begin{align} \label{defA}
A := \frac{1}{a} \sum_{k=1}^K         \sum_{\substack{i, i' \in T_k \\ i < i'}} (  x(i) -  x(i') )(  x(i) -  x(i') )^T \end{align}
We rewrite $A$ as $ \frac{1}{a} \sum_{k=1}^K    \dot{X_k}^T \dot{X_k} :=  \dot{X}^T \dot{X}$. Here $\dot{X} \in \mathbb{R}^{g \times d}$ is the matrix that have as row the difference between two vectors in $X$ that belongs to the same class. We can approximate the matrix $A$ by subsampling from all possible pairs $(x(i), x(i'))$ from each class. In our experiment we built $\dot{X}$ with a constant sample size of $g = 10^{4}$ derivatives from all the possible derivatives for a given class.  It is not hard to see that the weight vectors $w_j$ that correspond to the minima of equation (\ref{delta})  are the eigenvectors associated with the smallest eigenvalues of the generalized eigenvalue problem $AW=\Lambda BW$ \cite{friedman2001elements, sameh1982trace}, where  $\Lambda = Diag\left[ \lambda_1, ... \lambda_n \right] $ is the diagonal matrix of eigenvalues, and $W$ is the matrix of generalized eigenvectors.  Picking the $K-1$ smallest eigenvectors will allow us to create the $d \times (K-1)$ matrix $W$ that will project our data to the slow feature space. In other words, SFA reduces to a generalized eigenvalue problem.\\

Computationally, the training part of the SFA algorithm has two steps. First, the matrix $X$ is mapped to the matrix $Z = XB^{-1/2}$ of the whitened data. Whitening data matrix $X$ consist in diagonalizing the covariance matrix $X^TX$ of the dataset, and transforming the data such that the covariance matrix becomes the identity. Because of whitening, we reduced the generalized eigenvalue problem to a normal eigenvalue problem. Then, $Z$ is projected onto the space spanned by the eigenvectors associated to the $K-1$ smallest eigenvalues of the matrix $\dot{Z}^T\dot{Z}$. Here $\dot{Z}$ is defined similar to $\dot{X}$: by sampling the point-wise differences of the whitened data. Because of this, we can redefine the derivative covariance matrix as $A := \dot{Z}^T\dot{Z}=(B^{-1/2})^T \dot{X}^T \dot{X}B^{-1/2}$. \\


Remark that the best classical algorithms for operating on matrices and performing all the necessary linear algebraic procedures, are currently polynomial in the matrix dimensions. In fact, when the dimension of the input vectors becomes too large, procedures like full PCA or SFA become infeasible since their complexity is $\tilde{O}(min(n^2d, d^2n))$ \cite{ross2008incremental}.\\

\paragraph{Quantum algorithms for linear algebra}\label{quantum}
In what follows we adopt the convention that the matrices stored in QRAM are  pre-scaled, i.e. $\norm{M}_2 = 1$,  where $\norm{M}_2$ is the spectral norm (see \cite{kerenidis2017quantumsquares} for an efficient procedure for this normalization). In recent work \cite{CGJ18, GSLW18}, QRAM based procedures for matrix multiplication and inversion have been discovered with running time
$\tilde{O}(\kappa(M)\mu(M)\log(1/\epsilon))$, where $\kappa$ is the condition number of the matrix $M$, (i.e. the ratio between the biggest and smallest singular value), $\epsilon$ is the error, and the parameter $\mu(M)$ is defined as 
	$$\mu(M)= \min_{p \in P} (\norm{M}_F, \sqrt{s_{2p}(M)s_{2(1-p)}(M^{T})}),$$ 
	for $s_{p}(M) := \max_{i \in [n]} \norm{m_i}_p^p$ where $\norm{m_i}_p$ is the $\ell_p$ norm of the i-th row of $M$, and $P$ is a finite set of size $O(1) \in [0,1]$.
	Note that $\mu(M) \leq \norm{M}_{F} \leq \sqrt{d}$ as we have assumed that $\norm{M}_2\leq 1$. We will see that it will be convenient to choose $\mu$ to be the maximum $\ell_1$ norm of the rows of the matrix, i.e. $\mu(M)=\norm{M}_\infty$. In these works, the notion of block encoding is used, which is shown to be equivalent to our notion of having an efficient data structure as the one described in the Supplementary Metherial.  In order to get the polylogarithmic dependence on the precision parameter $\epsilon$, the authors show how to perform the linear algebra procedures directly through the technique of qubitization and without explicitly estimating coherently the singular values. Another important advantage of the new methods is that it provides easy ways to manipulate sums or products of matrices. We start by stating more precisely the results in \cite{CGJ18,GSLW18} about the quantum algorithms for linear algebraic procedures that we will use for proving the runtime and correctness of the QSFA algorithm.

\begin{theorem}[Matrix algebra \cite{CGJ18,GSLW18}]\label{matrix_algebra_improved}  Let $M := \sum_{i} \sigma_iu_iv_i^T \in \mathbb{R}^{d \times d}$ such that $\norm{M}_2 =1$, and a vector $x \in \mathbb{R}^d$ stored in QRAM. There exist quantum algorithms that with probability at least $1-1/poly(d)$ return
	\begin{enumerate}[(i)]
		\item a state $\ket{z}$ such that $| \ket{z} - \ket{Mx}| \leq \epsilon$ in time  $\tilde{O}(\kappa(M)\mu(M)\log(1/\epsilon))$  
		\item a state $\ket{z}$ such that $|\ket{z} - \ket{M^{-1}x}| \leq \epsilon$ in time $\tilde{O}(\kappa(M)\mu(M)\log(1/\epsilon))$
	\end{enumerate}
	One can also get estimates of the norms with multiplicative error $\eta$ by increasing the running time by a factor $1/\eta$.
\end{theorem}

\begin{theorem}[Matrix algebra on products of matrices \cite{CGJ18,GSLW18}]\label{matrix_algebra_product_improved}  Let $M_1, M_2 \in \mathbb{R}^{d \times d}$ such that $\norm{M_1}_2= \norm{M_2}_2 =1$, $M=M_1M_2$,
	and a vector $x \in \mathbb{R}^d$ stored in QRAM. There exist quantum algorithms that with probability at least $1-1/poly(d)$ return
	\begin{enumerate}[(i)]
		\item a state $\ket{z}$ such that $| \ket{z} - \ket{Mx}| \leq \epsilon$ in time $\tilde{O}(\kappa(M)(\mu(M_1)+\mu(M_2))\log(1/\epsilon))$  
		\item a state $\ket{z}$ such that $|\ket{z} - \ket{M^{-1}x}| \leq \epsilon$ in time $\tilde{O}(\kappa(M)(\mu(M_1)+\mu(M_2))\log(1/\epsilon))$  
		\item  a state $\ket{M_{\leq \theta, \delta}^+M_{\leq \theta, \delta}x}$  in time $\tilde{O}(\frac{ (\mu(M_1)+\mu(M_2)) \norm{x}}{\delta \theta \norm{M^{+}_{\leq \theta, \delta}M_{\leq \theta, \delta}x}})$ 
	\end{enumerate}
	One can also get estimates of the norms with multiplicative error $\eta$ by increasing the running time by a factor $1/\eta$.
\end{theorem}


\section{Quantum Slow Feature Analysis}\label{q_sfa}
Our goal is to devise a quantum algorithm that maps a quantum state corresponding to the data $X$ to a quantum state $Y$ that represent the data in the slow feature space of the SFA algorithm. Formally, $U_{QSFA}$ maps the state $\ket{X} :=  \frac{1}{\norm{X}_F} \sum_{i=1}^{n} \norm{x(i)} \ket{i}\ket{x(i)} $ to $\ket{Y} :=  \frac{1}{\norm{Y}_F} \sum_{i=0}^n\norm{y(i)}\ket{i}\ket{y(i)}$. As the classical algorithm, QSFA is divided in two parts. In the first step we whiten the data, i.e. we map the state $\ket{X}$ to the state $\ket{Z}$, and in the second step we project $\ket{Z}$ onto the subspace spanned by the smallest eigenvectors of the whitened derivative covariance matrix $A=\dot{Z}^T\dot{Z}$. As we know classically, that the data is whitened by multiplying $X$ with the matrix $B^{-1/2} = (X^TX)^{-1/2}$, in other words we build $\ket{Z}:=\ket{B^{-1/2}X}$. The results in \cite{CGJ18} state that the time to multiply with $B^{-1/2}$ is the same as the time to multiply with $X$ and hence we can perform this multiplication using Theorem \ref{matrix_algebra_improved}. Now we want to project this state onto the subspace spanned by the eigenvectors associated to the $K-1$ ``slowest'' eigenvectors (i.e. associated with the smallest singular values) of the whitened derivative covariance matrix $A :=\dot{Z}^T\dot{Z}$, where
$\dot{Z}$ is the whitened derivative matrix $\dot{Z} = \dot{X}B^{-1/2}$. The procedure follows the projection procedure that was used in \cite{kerenidis2016recommendation} for recommendation systems, albeit the projection is now in the lowest eigenspectrum instead of the highest one. 
Let $\theta$ a threshold value and $\delta$ a precision parameter. 
With $A_{\leq \theta, \delta}$ we denote a projection of the matrix $A$ onto the vector subspace spanned by the union of the singular vectors associated to singular values that are smaller than $\theta$ and some subset of singular vectors whose corresponding singular values are in the interval $[\theta, (1+\delta) \theta]$. 
Again, we note that the eigenvalues of $A$ are the squares of the singular values of $\dot{Z}$, and the two matrices share the same row space: $\dot{Z} = U\Sigma V^T $, and $A = V\Sigma^2 V^T$. Note also that whitening the derivatives is equal to taking the derivatives of the whitened data. We can therefore use the quantum algorithms for linear algebra and state the following. 

\begin{theorem}[QSFA]\label{theorem_qsfa}
	Let $X = \sum_i \sigma_i u_iv_i^T \in \mathbb{R}^{n\times d}$ and its derivative matrix
	$\dot{X} \in \mathbb{R}^{n \log n \times d}$ stored in QRAM as described in the Supplementary material. Let $\epsilon, \theta, \delta, \eta >0$.
	There exists a quantum algorithm that produces as output a state $\ket{\overline{Y}}$ with 
	$| \ket{\overline{Y}} - \ket{A^+_{\leq \theta, \delta}A_{\leq \theta, \delta} Z} | \leq \epsilon$ 
	in time 
	\begin{align}\tilde{O} \Big(  \big( \kappa(X)\mu(X)\log (1/\varepsilon) + \frac{  ( \mu({X})+ \mu(\dot{X}) ) }{\delta\theta} \big)
	\nonumber \\ \times \frac{|norm{Z}}{ \norm{A^+_{\leq \theta, \delta}A_{\leq \theta, \delta} {Z} }} \Big) \end{align}
	and an estimator $\overline{\norm{Y}}$ with $| \overline{\norm{Y}} - \norm{Y} | \leq \eta \norm{Y}$
	with an additional $1/\eta$ factor in the running time.
\end{theorem}

A priori, one does not know the appropriate threshold value $\theta$, but this can be found efficiently using binary search. Note also that the output of the algorithm is not a classical description of the projected inputs but a quantum state that corresponds to these projected inputs, but this is sufficient for using the classification procedure described afterwards. If necessary, we can recover classically the $K-1$ slow feature vectors using process tomography. We present the proof Theorem \ref{theorem_qsfa} in the Supplementary Material section. 

\section{Quantum Frobenius Distance classifier}\label{q_classification}


To perform the classification of the MNIST dataset we propose the Quantum Frobenius Distance classifier (QFD), a quantum classification algorithm designed with the ease of implementation in mind. The QFD assigns a test point $x(0)$ to the cluster $k$ whose points have minimum normalized average squared $\ell_2$ distance to $x(0)$. Let $X_k$ be defined as the matrix whose rows are the vectors corresponding to the $k$-th cluster, and $|T_k|$ is the number of elements in that cluster. 
For the test point $x(0)$ and a class $k$, define the matrix $X(0) \in \mathbb{R}^{|T_k| \times d}$ which just repeats the row $x(0)$ $|T_k|$ times. Then, we define 
\begin{equation}\label{frobdist}
    F_k( x(0)) = \frac{ \norm{X_k - X(0)}_F^2}{2 ( \norm{X_k}_F^2+ \norm{X(0)}_F^2) }
\end{equation}
as the average normalized squared distance between $x(0)$ and the cluster $k$. 
Let $h : \mathcal{X} \to [K]$ our classification function. Then, the hypothesis of the class made by the QFD classifier on $x(0)$ is:
\begin{align}
h(x(0)) :=  \argmin_{k \in [K]} F_k( x(0))
\end{align}
Our algorithm - which is described in detail in the Supplementary Material section, estimates $F_k(x(0))$ efficiently, assuming the vectors and their norms are loaded into the quantum computer either from the QRAM, or come directly from some quantum process. For its implementation we use little more than the ability to query the QRAM and an Hadamard gate. In fact, we first create the following state by using one control qubit and the ability to efficiently create the quantum states corresponding to the data points

\begin{align}\label{qfd}   \frac{1}{\sqrt{N_k}} \Big( \ket{0}  \sum_{i \in T_k}  \norm{x(0)} \ket{i} \ket{x(0)}+ \nonumber \\\ket{1} \sum_{i \in T_k}  \norm{x(i)} \ket{i}\ket{x(i)} \Big)
\end{align}


Then, we apply a Hadamard on the leftmost qubit, and note that the probability of measuring $1$ is exactly equal to $F_k(x(0))$. We can estimate the probability of $1$ by measuring in the computational basis in time $\tilde{O}(\frac{1}{\eta^2})$, or in  $\tilde{O}(\frac{1}{\eta})$ using amplitude amplification. We will see that in fact $\eta$ does not have to be very small in order to classify correctly the MNIST dataset after the dimensionality reduction (we can take $\eta=1/10$), since the clusters are pretty well separated. The running time of this procedure is $\tilde{O}(\frac{K}{\eta})$, when we assume that the data is stored in QRAM or it can be efficiently created by a quantum procedure.

\section{The quantum classifier}
For each cluster $k$, we first use the QSFA procedure to project the input onto the slow feature space, where we use the QFD classifier. More precisely, for each class we use the QSFA procedure that maps $\ket{X_k}$ to a state $\ket{\overline{Y_k}}$, which is $\varepsilon$-close to the state $\ket{Y_k} = \frac{1}{\norm{Y_k}_F}\sum_{i \in T_k} \norm{y(i)}\ket{i}\ket{y(i)} $. We can also use the same procedure to construct the vector $\ket{y(0)}$ and hence construct a state equivalent to the state in Eq. \ref{qfd} but in the slow feature space. Then, the QFD algorithm assigns a class to $y(0)$ by finding the closest cluster with respect to the Frobenius distance in Eq. \ref{frobdist}.

\paragraph*{Analysis of accuracy on the MNIST data set.} 
Since currently available quantum hardware cannot run our algorithm or any significant part of it, we simulated QSFA and QFD on an Atos QLM with 6 TB of RAM. We simulated our combined quantum classifier on the MNIST dataset of handwritten digits, a standard dataset used to benchmark machine learning algorithms. Details on the simulations can be found in the Supplementary Material section. Our software heavily relied on python packages for numerical computation and machine learning: scikit-learn \cite{scikit-learn} and the scipy ecosystem \cite{scipy}. To gauge the accuracy of our classification procedure, we first need to take into account the inherent errors in the quantum procedures (for example in estimating the singular values), and second, we need to test our classification algorithm, since our notion of distance is not among the ones used classically. Then, we provide some estimation of the running time of the quantum algorithm, by computing the parameters that appear in the asymptotic complexity of the algorithm. Our estimation does not include terms that depend on the hardware implementations of the quantum algorithm, that are for the moment not clear. For example, the fact that a quantum step may be slower than a classical step, or the possible overhead due to the error correcting codes. Our goal here is to see if, a priori, the quantum algorithm itself is more efficient than the classical one. If there is no speedup in our analysis, then one cannot hope to see a speedup in practice. If there is a significant speedup, then there is hope that part of this speedup (and which part depends on how good the whole technological stack will be, like hardoware and error correcting codes) can be seen in practice.  \\


The MNIST dataset, first studied in \cite{lecun1998gradient,Qiao2007}, is composed of $n=60000$ grayscale images of handwritten digits, $28 \times 28$ pixels. We will follow closely the methodoloy used to apply classical SFA to the MNIST dataset, detailed in \cite{Berkes2005pattern}. The methodology is as follows: first, the dimension of the dataset is reduced with a PCA to something like 35 (or around 90, depending on the polynomial expansion degree we use in the following step). Second, a polynomial expansion of degree 2 or 3 is applied, hence making the dimension up to $10^4$. For example, with polynomial expansion of degree 2 we mean that a vector $(x_1, x_2, \ldots, x_d)$ is mapped to  $(x_1, \ldots, x_d, x_1^2, x_1x_2,\ldots, x_d^2)$. Third, the data is scaled to zero mean and unit variance. We tested our algorithm while increasing the dimension of the initial PCA on the training set, hoping the accuracy will get even better when we increase the dimension of the input.

\begin{figure}
	\centering 
	\includegraphics[width=0.5\textwidth]{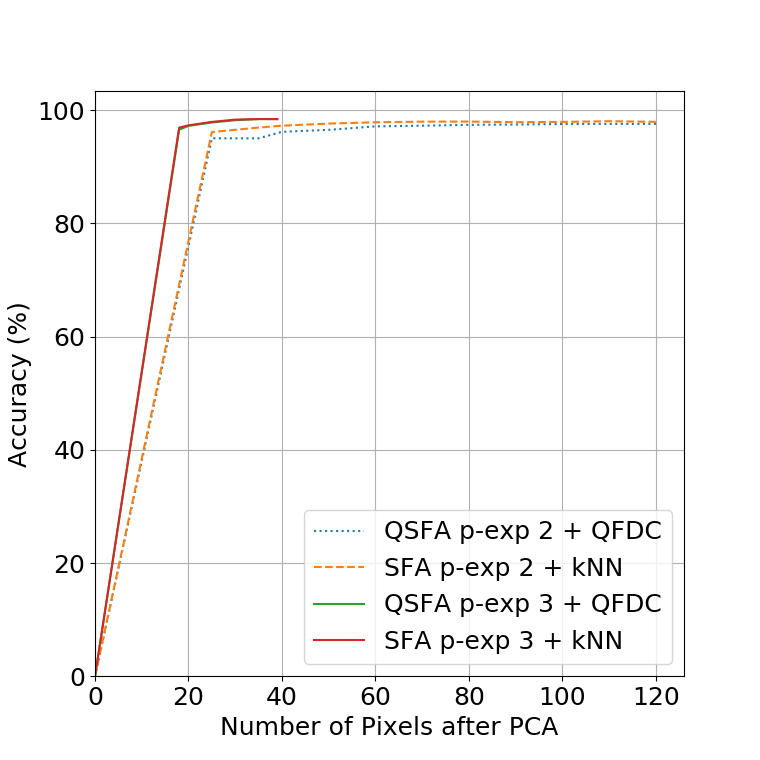} 
	\caption{The accuracy of our quantum classifier (QSFA with QFDC) versus the classical classifier (SFA with k-NearestNeighbor) for polynomial expansion 2 and 3, using different number of pixels via PCA. For polynomial expansion 2, our quantum classifier (bottom dotted line) reaches the accuracy of the classical one (dashed line) for large enough dimensions. For polynomial expansion 3 (upper thick line), the two are almost indistinguishable. In this experiment, the error for the quantum classifier on the quantum linear algebra subroutines is $\epsilon=10^{-5}$ and $\kappa_t$ is less than $200$.}
	\label{QSFA_compare}
\end{figure}




In Figure \ref{QSFA_compare} we plot the accuracy (percentage of correctly classified digits from the test set) of the quantum classifier (QSFA + QFD) using polynomial expansion of degree 2 and 3, showing that we achieve very good accuracy. On the $x$ axis we changed the initial resolution of the image using PCA. The highest performance accuracy was $98.5\%$ with polynomial expansion of degree $3$ and PCA dimension of $36$, which took about an hour of simulation. Given the favorable dependence on the dimension, we expect a quantum computer to achieve even higher accuracy given a polynomial expansion of a higher degree, and more data.\\




We also tested the accuracy of our classifier when instead of the condition number $\kappa$ we use a condition threshold $\kappa_t$ (i.e. we discarded singular values of the matrices below a certain threshold, which we vary between 30 and 200). This usually improves the running time by a factor of 2 while keeping the accuracy unchanged. The threshold of the condition number of $X$ is chosen to retain 99.5\% of the singular values. In this case, the accuracy for polynomial expansion 2 remains practically unchanged, while for polynomial expansion 3 there is a small decline, with the maximum becoming $98.2\%$ for PCA dimension 30. We can boost again the accuracy in the following way: instead of running the classifier once, we run it a few times (around 10) and do a majority vote before labelling the new data point. This increases the accuracy of polynomial expansion 3 a bit, for example for PCA 36, the accuracy goes from $97.7\%$ to $97.9\%$. Note that while we tried to optimize the parameters in order to achieve the best possible accuracy, it is very probable there are even better parameters that in practice can be found by a hyper-parameter tuning algorithm. This could further increase the accuracy of the quantum classifier.\\

\paragraph*{Running time estimation.} The asymptotic running time of the quantum classifier is given by the running time of the QSFA times a factor $K/\eta^2$ (or $K/\eta$ with amplitude amplification) that comes from the QFD. For the MNIST dataset, we estimated the value of the parameters that affects the runtime of the quantum classifier's (training and testing part together). For instance, we measured the condition number of the matrices used in the quantum linear algebra operations. We also estimated the tolerable error in $\epsilon, \delta, \theta, $ and $\eta$, such that the generalization error is comparable to classical algorithms. This has been possible by perturbing the model with some kind of noise, as described in the Supplemental Material. Overall, putting the orders of all parameters together, the estimated asymptotic running time  is of the order of $10^7$. Again, this is just an estimate of the steps of the quantum algorithm. More importantly, the behaviour of the parameters as the dimension increases, give further evidence that the quantum classifier can be more efficient that a classical classifier, whose running time is of the order of $10^{13}$ for the same input dimension. Moreover, the fact that all parameters that appear in the running time of the quantum classifier seem to increase very slowly (or not at all) as we increase the number and dimension of the data points, leads us to believe that one could still have an efficient quantum classifier with much higher number and dimension of points, thus eventually providing much higher classification accuracy.

\section{Discussion}
We provided evidence that quantum computers with quantum access to data can be useful for solving real-world problems by designing an efficient classifier. We achieved accuracy comparable to the best classical algorithms, with improved running time. It would be interesting to see if the same performance is achievable also on different datasets, which we do not see any reason why not. \\

In one of the experiments our quantum classifier performs the training and testing part together, i.e. one composes the QSFA and QFD procedures, in contrast to the classical case where one first uses the training to obtain the model classically and then uses the model for the testing. On one hand, this allows us to quickly classify new data without having to wait for an extensive training period. In fact, according to our rough estimates, one can classify $10^6$ images, before the classical algorithm finishes its training part. On the other hand, if the testing part contains a very large number of data points (many orders of magnitude more than the training part), then one might want to use the quantum procedure to extract the classical SFA model, in other words, find the matrix $W$, such that $Y=XW$. For this, it suffices to use the QSFA algorithm with initial state the totally mixed state, in which case the quantum output of the QSFA procedure is in fact the state that corresponds to the matrix $W$. Performing efficient tomography for this state, using the procedure in \cite{kerenidis2018quantum}, we can accurately find the matrix $W$ and store it in QRAM. Then, the testing can be done classically in time $O(Kd)$.\\

We envision the utility of our classifier in conjunction with other quantum machine learning algorithms. For instance, Projective Simulation \cite{melnikov2017projective} is a Reinforcement Learning algorithm based on a random walk over a graph. The graph represents the memory of an agent that acts on a certain environment. The random walk starts from a node (or superposition of nodes) decided by an input-coupling function. QSFA could be used to treat and pre-process high-dimensional input signals in agents that use Projective Simulation as input-coupling function of external stimuli in the memory model of the agent. This would resemble even further what we currently believe to be the architecture of the brain, where SFA is used to model complex cells in the primary visual cortex V1: the first cortical area dedicated to visual processing \cite{berkes2005rich}, and Projective Simulation is used to model high level cognitive functions which emerges from a model  of an episodic and compositional memory for the agent (so to model creativity, curiosity, and so on..)\cite{dunjko2018machine}. Note also, that the quantum algorithm for SFA can also be used for classification via Fisher Linear Discriminant \cite{klampfl2009replacing} and Laplacian Eigenmaps \cite{sprekeler2011relation}.\\


Note also that it might be more important not to see our quantum classifier as a faster algorithm but as a way to increase the accuracy. As we said, the training stage is limited by the dimension of the input and the fact that our quantum classifier depends only polylogarithmically on the dimension and all other parameters remain stable when the dimension increases, will enable us to use much higher dimension in the training stage, thus hopefully improving the accuracy of the classifier. Despite being an algorithm based on linear algebra, techniques like polynomial expansions can help in capture nonlinearities hidden in the data, thus improving further the accuracy of the classifier. Alas, this comes at the cost of increasing the dimension of the dataset, and thus an increase of the runtime of classical algorithms. With quantum algorithms with polylogarithmic dependence in the dimension of the dataset we believe these techniques can be even more beneficial.

\section*{Acknowledgement}
We thank Anupam Prakash and Andr{\'a}s Gily{\'e}n for helpful discussions. This research was supported by QuantAlgo, Quantex, QuData and ANR. The experiments have been performed on a Atos QLM. 
	
\section*{Author contribution}
Both authors contributed equally to the theoretical part of the work. Luongo made the simulations of the quantum algorithms.

\section*{Competing interests}
The authors declare that there are no competing interests.

\newpage
\bibliographystyle{alpha}	
\bibliography{Mendeley,further_bibliography}

\newcommand{\etalchar}[1]{$^{#1}$}
\begin{thebibliography}{BRGBPO17}

\bibitem[ADBL19]{arrazola2019quantum}
Juan~Miguel Arrazola, Alain Delgado, Bhaskar~Roy Bardhan, and Seth Lloyd.
\newblock Quantum-inspired algorithms in practice.
\newblock {\em arXiv preprint arXiv:1905.10415}, 2019.

\bibitem[BDLK19]{bang2019optimal}
Jeongho Bang, Arijit Dutta, Seung-Woo Lee, and Jaewan Kim.
\newblock Optimal usage of quantum random access memory in quantum machine
  learning.
\newblock {\em Physical Review A}, 99(1):012326, 2019.

\bibitem[Ber05]{Berkes2005pattern}
Pietro Berkes.
\newblock {Pattern Recognition with Slow Feature Analysis}.
\newblock {\em Cognitive Sciences EPrint Archive (CogPrints)}, 4104, 2005.

\bibitem[BLS19]{benedetti2019parameterized}
Marcello Benedetti, Erika Lloyd, and Stefan Sack.
\newblock Parameterized quantum circuits as machine learning models.
\newblock {\em arXiv preprint arXiv:1906.07682}, 2019.

\bibitem[BRGBPO17]{Benedetti2016assistedgraphicalmodel}
Marcello Benedetti, John Realpe-G{\'o}mez, Rupak Biswas, and Alejandro
  Perdomo-Ortiz.
\newblock Quantum-assisted learning of hardware-embedded probabilistic
  graphical models.
\newblock {\em Physical Review X}, 7(4):041052, 2017.

\bibitem[BW04]{blaschke2004independent}
Tobias Blaschke and Laurenz Wiskott.
\newblock Independent slow feature analysis and nonlinear blind source
  separation.
\newblock In {\em International Conference on Independent Component Analysis
  and Signal Separation}, pages 742--749. Springer, 2004.

\bibitem[BW05]{berkes2005rich}
P.~Berkes and L.~Wiskott.
\newblock {Slow feature analysis yields a rich repertoire of complex cell
  properties}.
\newblock {\em Journal of Vision}, 5(6):9--9, 2005.

\bibitem[CD16]{iris2016discriminant}
Iris Cong and Luming Duan.
\newblock Quantum discriminant analysis for dimensionality reduction and
  classification.
\newblock {\em New Journal of Physics}, 18(7):073011, 2016.

\bibitem[CGJ19]{CGJ18}
Shantanav Chakraborty, Andr{\'a}s Gily{\'e}n, and Stacey Jeffery.
\newblock The power of block-encoded matrix powers: Improved regression
  techniques via faster hamiltonian simulation.
\newblock In {\em 46th International Colloquium on Automata, Languages, and
  Programming (ICALP 2019)}. Schloss Dagstuhl-Leibniz-Zentrum fuer Informatik,
  2019.

\bibitem[CW12]{childs2012hamiltonian}
Andrew~M Childs and Nathan Wiebe.
\newblock Hamiltonian simulation using linear combinations of unitary
  operations.
\newblock {\em arXiv preprint arXiv:1202.5822}, 2012.

\bibitem[DB18]{dunjko2018machine}
Vedran Dunjko and Hans~J Briegel.
\newblock Machine learning \& artificial intelligence in the quantum domain: a
  review of recent progress.
\newblock {\em Reports on Progress in Physics}, 81(7):074001, 2018.

\bibitem[EBW12]{escalante2012slow}
Alberto~N Escalante-B and Laurenz Wiskott.
\newblock Slow feature analysis: Perspectives for technical applications of a
  versatile learning algorithm.
\newblock {\em KI-K{\"u}nstliche Intelligenz}, 26(4):341--348, 2012.

\bibitem[FKV04]{frieze2004fast}
Alan Frieze, Ravi Kannan, and Santosh Vempala.
\newblock Fast monte-carlo algorithms for finding low-rank approximations.
\newblock {\em Journal of the ACM (JACM)}, 51(6):1025--1041, 2004.

\bibitem[FN18]{farhi2018classification}
Edward Farhi and Hartmut Neven.
\newblock Classification with quantum neural networks on near term processors.
\newblock {\em arXiv preprint arXiv:1802.06002}, 2018.

\bibitem[GLW13]{Gu2013supervised}
Xingjian Gu, Chuancai Liu, and Sheng Wang.
\newblock Supervised slow feature analysis for face recognition.
\newblock {\em Biometric Recognition}, page 178, 2013.

\bibitem[GSLW19]{GSLW18}
Andr{\'a}s Gily{\'e}n, Yuan Su, Guang~Hao Low, and Nathan Wiebe.
\newblock Quantum singular value transformation and beyond: exponential
  improvements for quantum matrix arithmetics.
\newblock In {\em Proceedings of the 51st Annual ACM SIGACT Symposium on Theory
  of Computing}, pages 193--204, 2019.

\bibitem[GWOB19]{grant2019initialization}
Edward Grant, Leonard Wossnig, Mateusz Ostaszewski, and Marcello Benedetti.
\newblock An initialization strategy for addressing barren plateaus in
  parametrized quantum circuits.
\newblock {\em arXiv preprint arXiv:1903.05076}, 2019.

\bibitem[HHL09]{HarrowHassidim2009HHL}
Aram~W. Harrow, Avinatan Hassidim, and Seth Lloyd.
\newblock {Quantum Algorithm for Linear Systems of Equations}.
\newblock {\em Physical Review Letters}, 103(15):150502, 10 2009.

\bibitem[HTF09]{friedman2001elements}
Trevor Hastie, Robert Tibshirani, and Jerome Friedman.
\newblock {\em {The Elements of Statistical Learning}}, volume~1 of {\em
  Springer Series in Statistics}.
\newblock Springer New York, New York, NY, 2009.

\bibitem[HZZ{\etalchar{+}}19]{hann2019hardware}
Connor~T Hann, Chang-Ling Zou, Yaxing Zhang, Yiwen Chu, Robert~J Schoelkopf,
  Steven~M Girvin, and Liang Jiang.
\newblock Hardware-efficient quantum random access memory with hybrid quantum
  acoustic systems.
\newblock {\em arXiv preprint arXiv:1906.11340}, 2019.

\bibitem[JOP{\etalchar{+}}18]{scipy}
Eric Jones, Travis Oliphant, Pearu Peterson, et~al.
\newblock {SciPy}: Open source scientific tools for {Python}, 2001-2018.
\newblock [Online; accessed 2018-05-19].

\bibitem[JPC{\etalchar{+}}19]{jiang2019experimental}
N~Jiang, Y-F Pu, W~Chang, C~Li, S~Zhang, and L-M Duan.
\newblock Experimental realization of 105-qubit random access quantum memory.
\newblock {\em npj Quantum Information}, 5(1):28, 2019.

\bibitem[Kay95]{kaynak1995methods}
C~Kaynak.
\newblock {\em {Methods of combining multiple classifiers and their
  applications to handwritten digit recognition}}.
\newblock PhD thesis, Institute of Graduate Studies in Science and Engineering,
  Bogazici University, 1995.

\bibitem[KM09]{klampfl2009replacing}
Stefan Klampfl and Wolfgang Maass.
\newblock {Replacing supervised classification learning by Slow Feature
  Analysis in spiking neural networks}.
\newblock In {\em Advances in Neural Information Processing Systems}, pages
  988--996, 2009.

\bibitem[KP17a]{kerenidis2017quantumsquares}
Iordanis Kerenidis and Anupam Prakash.
\newblock Quantum gradient descent for linear systems and least squares.
\newblock {\em arXiv preprint arXiv:1704.04992}, 2017.

\bibitem[KP17b]{kerenidis2016recommendation}
Iordanis Kerenidis and Anupam Prakash.
\newblock Quantum recommendation systems.
\newblock In {\em 8th Innovations in Theoretical Computer Science Conference
  (ITCS 2017)}. Schloss Dagstuhl-Leibniz-Zentrum fuer Informatik, 2017.

\bibitem[KP18]{kerenidis2018quantum}
Iordanis Kerenidis and Anupam Prakash.
\newblock A quantum interior point method for lps and sdps.
\newblock {\em arXiv preprint arXiv:1808.09266}, 2018.

\bibitem[LBBH98]{lecun1998gradient}
Yann LeCun, L{\'e}on Bottou, Yoshua Bengio, and Patrick Haffner.
\newblock Gradient-based learning applied to document recognition.
\newblock {\em Proceedings of the IEEE}, 86(11):2278--2324, 1998.

\bibitem[LC10]{Qiao2007}
Yann LeCun and Corinna Cortes.
\newblock {MNIST} handwritten digit database.
\newblock 2010.

\bibitem[LeC98]{lecun1998mnist}
Yann LeCun.
\newblock The mnist database of handwritten digits.
\newblock {\em http://yann. lecun. com/exdb/mnist/}, 19998.

\bibitem[LGZ16]{lloyd2016quantumtopological}
Seth Lloyd, Silvano Garnerone, and Paolo Zanardi.
\newblock {Quantum algorithms for topological and geometric analysis of data}.
\newblock {\em Nature communications}, 7:10138, 2016.

\bibitem[LMR13]{Lloyd2013}
Seth Lloyd, Masoud Mohseni, and Patrick Rebentrost.
\newblock {Quantum principal component analysis}.
\newblock {\em Nature Physics}, 10(9):631--633, 7 2013.

\bibitem[LR18]{liu2018quantum}
Nana Liu and Patrick Rebentrost.
\newblock Quantum machine learning for quantum anomaly detection.
\newblock {\em Physical Review A}, 97(4):042315, 2018.

\bibitem[Mac02]{mackay2003information}
David~JC MacKay.
\newblock {\em {Information theory, inference and learning algorithms}}.
\newblock Cambridge University Press, 2002.

\bibitem[MBS{\etalchar{+}}18]{mcclean2018barren}
Jarrod~R McClean, Sergio Boixo, Vadim~N Smelyanskiy, Ryan Babbush, and Hartmut
  Neven.
\newblock Barren plateaus in quantum neural network training landscapes.
\newblock {\em arXiv preprint arXiv:1803.11173}, 2018.

\bibitem[MMDB17]{melnikov2017projective}
Alexey~A Melnikov, Adi Makmal, Vedran Dunjko, and Hans~J Briegel.
\newblock Projective simulation with generalization.
\newblock {\em Scientific reports}, 7(1):14430, 2017.

\bibitem[OMA{\etalchar{+}}17]{otterbach2017unsupervised}
JS~Otterbach, R~Manenti, N~Alidoust, A~Bestwick, M~Block, B~Bloom, S~Caldwell,
  N~Didier, E~Schuyler Fried, S~Hong, et~al.
\newblock Unsupervised machine learning on a hybrid quantum computer.
\newblock {\em arXiv preprint arXiv:1712.05771}, 2017.

\bibitem[PPR19]{park2019circuit}
Daniel~K Park, Francesco Petruccione, and June-Koo~Kevin Rhee.
\newblock Circuit-based quantum random access memory for classical data.
\newblock {\em Scientific reports}, 9(1):3949, 2019.

\bibitem[Pra14]{Prakash:EECS-2014-211}
Anupam Prakash.
\newblock {\em Quantum Algorithms for Linear Algebra and Machine Learning.}
\newblock PhD thesis, EECS Department, University of California, Berkeley, Dec
  2014.

\bibitem[PVG{\etalchar{+}}11]{scikit-learn}
F~Pedregosa, G~Varoquaux, A~Gramfort, V~Michel, B~Thirion, O~Grisel, M~Blondel,
  P~Prettenhofer, R~Weiss, V~Dubourg, J~Vanderplas, A~Passos, D~Cournapeau,
  M~Brucher, M~Perrot, and E~Duchesnay.
\newblock {Scikit-learn: Machine Learning in {\{}P{\}}ython}.
\newblock {\em Journal of Machine Learning Research}, 12:2825--2830, 2011.

\bibitem[RL18]{rebentrost2018quantum}
Patrick Rebentrost and Seth Lloyd.
\newblock Quantum computational finance: quantum algorithm for portfolio
  optimization.
\newblock {\em arXiv preprint arXiv:1811.03975}, 2018.

\bibitem[RLLY08]{ross2008incremental}
David~A Ross, Jongwoo Lim, Ruei-Sung Lin, and Ming-Hsuan Yang.
\newblock Incremental learning for robust visual tracking.
\newblock {\em International journal of computer vision}, 77(1-3):125--141,
  2008.

\bibitem[RML14]{rebentrost2014quantumsvm}
Patrick Rebentrost, Masoud Mohseni, and Seth Lloyd.
\newblock {Quantum support vector machine for big data classification}.
\newblock {\em Physical review letters}, 113(13):130503, 7 2014.

\bibitem[SBSW18]{webie2018quantumcentric}
Maria Schuld, Alex Bocharov, Krysta Svore, and Nathan Wiebe.
\newblock Circuit-centric quantum classifiers.
\newblock {\em arXiv preprint arXiv:1804.00633}, 2018.

\bibitem[SFP17]{Schuld2017interference}
Maria Schuld, Mark Fingerhuth, and Francesco Petruccione.
\newblock Implementing a distance-based classifier with a quantum interference
  circuit.
\newblock {\em EPL (Europhysics Letters)}, 119(6):60002, 2017.

\bibitem[SJC{\etalchar{+}}14]{sun2014deepsfaactionrecognition}
Lin Sun, Kui Jia, Tsung-Han Chan, Yuqiang Fang, Gang Wang, and Shuicheng Yan.
\newblock Dl-sfa: deeply-learned slow feature analysis for action recognition.
\newblock In {\em Proceedings of the IEEE Conference on Computer Vision and
  Pattern Recognition}, pages 2625--2632, 2014.

\bibitem[Spr11]{sprekeler2011relation}
Henning Sprekeler.
\newblock {On the Relation of Slow Feature Analysis and Laplacian Eigenmaps}.
\newblock {\em Neural Computation}, 23(12):3287--3302, 2011.

\bibitem[SW82]{sameh1982trace}
Ahmed~H Sameh and John~A Wisniewski.
\newblock {A trace minimization algorithm for the generalized eigenvalue
  problem}.
\newblock {\em SIAM Journal on Numerical Analysis}, 19(6):1243--1259, 1982.

\bibitem[SW08]{Sprekeler2008understandingframework}
Henning Sprekeler and Laurenz Wiskott.
\newblock {Understanding Slow Feature Analysis: A Mathematical Framework}.
\newblock {\em Cognitive Sciences EPrint Archive (CogPrints)}, 6223, 2008.

\bibitem[SZW14]{sprekeler2014extension}
Henning Sprekeler, Tiziano Zito, and Laurenz Wiskott.
\newblock {An extension of slow feature analysis for nonlinear blind source
  separation.}
\newblock {\em Journal of machine learning research}, 15(1):921--947, 2014.

\bibitem[Tan18]{tang2018quantum}
Ewin Tang.
\newblock Quantum-inspired classical algorithms for principal component
  analysis and supervised clustering.
\newblock {\em arXiv preprint arXiv:1811.00414}, 2018.

\bibitem[Tan19]{tang2018quantum2}
Ewin Tang.
\newblock A quantum-inspired classical algorithm for recommendation systems.
\newblock In {\em Proceedings of the 51st Annual ACM SIGACT Symposium on Theory
  of Computing}, pages 217--228, 2019.

\bibitem[VBB17]{verdon2017quantum}
Guillaume Verdon, Michael Broughton, and Jacob Biamonte.
\newblock A quantum algorithm to train neural networks using low-depth
  circuits.
\newblock {\em arXiv preprint arXiv:1712.05304}, 2017.

\bibitem[WBF{\etalchar{+}}11]{scholarpedia2017SFA}
L.~Wiskott, P.~Berkes, M.~Franzius, H.~Sprekeler, and N.~Wilbert.
\newblock {S}low feature analysis.
\newblock {\em Scholarpedia}, 6(4):5282, 2011.
\newblock revision \#137965.

\bibitem[WBL12]{wiebe2012quantum}
Nathan Wiebe, Daniel Braun, and Seth Lloyd.
\newblock {Quantum Algorithm for Data Fitting}.
\newblock {\em Physical Review Letters}, 109(5):050505, 8 2012.

\bibitem[WW99]{wiskott1999learning}
{Wiskott Laurenz} and Laurenz Wiskott.
\newblock {Learning invariance manifolds}.
\newblock {\em Neurocomputing}, 26-27:925--932, 1999.

\bibitem[ZD12]{zhang2012slow}
{Zhang Zhang} and {Dacheng Tao}.
\newblock {Slow Feature Analysis for Human Action Recognition}.
\newblock {\em IEEE Transactions on Pattern Analysis and Machine Intelligence},
  34(3):436--450, 3 2012.

\end{thebibliography}

\appendix
\section{Quantum Slow Feature Analysis.}
In this section we provide the details of the QSFA algorithm introduced in Theorem 3 of the main text.

QSFA consists of two steps. The first step is the whitening, which can be performed in time $\tilde{O}(\kappa(X)\mu(X)\log(1/\epsilon))$ and provide the state $\ket{\overline{Z}}$. It is simple to verify that creating a state $\ket{Z}$ of whitened data such that $Z^TZ = I$ can be done using quantum access just to the matrix $X$, as $Z=XB^{-1/2}$. The second step is the projection of whitened data in the slow feature space, which is spanned by the eigenvectors of $A=\dot{Z}^T\dot{Z}$. This matrix shares the same right eigenvectors of $\dot{X}B^{-1/2}$, which is simple to check that we can efficiently access using the QRAM constructions of $X$ and $\dot{X}$. Using the algorithm for quantum linear algebra, we know that the projection (without the amplitude amplification) takes time equal to the ratio $\mu(X) +\mu(\dot{X})$ over the threshold parameter, in other words it takes time $\tilde{O}( \frac{(\mu({X})+ \mu(\dot{X}) }{\delta \theta}) $. Finally, the amplitude amplification and estimation depends on the size of the projection of $\ket{\overline{Z}}$ onto the slow eigenspace of $A$, more precisely it corresponds to the factor $\tilde{O}(\frac{\norm{\overline{Z}}}{ \norm{A^+_{\leq \theta, \kappa}A_{\leq \theta, \kappa} \overline{Z} }})$. This term is roughly the same if we look at $Z$ instead of $\overline{Z}$. Note also that $Z$ is the whitened data, which means that each whitened vector should look roughly the same on each direction. This implies that the ratio should be proportional to the ratio of the dimension of the whitened data over the dimension of the output signal. Note that in the case of the MNIST dataset this ratio is small enough. 

\begin{algorithm}[H]
	\floatname{algorithm}{Algorithm}
	\caption{Quantum Slow Feature Analysis}
	\label{alg_QSFA}
	\begin{algorithmic}[1]
	\Require
				\Statex Matrices $X \in \mathbb{R}^{n \times d}$ and $\dot{X} \in \mathbb{R}^{n \times d}$ in QRAM, parameters $\epsilon, \theta,\delta,\eta >0$. 
			\Ensure
				\Statex A state $\ket{\overline{Y}}$ such that $ | \ket{Y} - \ket{\overline{Y}} | \leq \epsilon$, with $Y = A^+_{\leq \theta, \delta}A_{\leq \theta, \delta} Z$
				\Statex
	
	\State Create the state $$\ket{X} :=  \frac{1}{\norm{X}_F} \sum_{i=1}^{n} \norm{x(i)} \ket{i}\ket{x(i)} $$
	
	\State (Whitening algorithm) Map $\ket{X}$ to $\ket{\overline{Z}}$ with $| \ket{\overline{Z}}  - \ket{Z} | \leq \epsilon $ and $Z=XB^{-1/2}.$
	
	
	\State (Projection in slow feature space) Project $\ket{\overline{Z}}$ onto the slow eigenspace of $A$ using threshold $\theta$ and precision $\delta$ (i.e. $A^+_{\leq \theta, \delta}A_{\leq \theta, \delta}\overline{Z}$) 
	
	\State Perform amplitude amplification and estimation on the register $\ket{0}$ with the unitary $U$ implementing steps 1 to 3, to obtain 
	$\ket{\overline{Y}}$ with $| \ket{\overline{Y}} - \ket{Y}  | \leq \epsilon $ and an estimate $\overline{\norm{Y}}$  with multiplicative error $\eta$.
	\end{algorithmic}
	\end{algorithm}


\section{The simulation}
The MNIST dataset, is a commonly used benchmark to test the validity of newly proposed classifiers. Classical classification techniques can achieve around 98-99\% accuracy, with neural network solutions exceeding 99\% (the MNIST dataset is quite simple and this is why it is often used as a first benchmark). As previously introduced in the main text, classical SFA has also been applied on the same dataset with accuracy 98.5\%, with initial PCA of dimension 35 and polynomial expansion of degree 3, and we will closely follow that classification procedure. Our goal is to study a quantum classifier with two properties: very good accuracy and efficiency. We detail our quantum classifier by going through the three parts in any classical classifier: preprocessing, training, and testing.
The error in the whitening procedure has been simulated by adding noise from a truncated Gaussian distribution centered on each singular value with unit variance. For the error in the projection part, this comes only from potentially projecting on a different space than the one wanted. By taking the right $\theta$ (around $0.3$ or $0.05$ depending on the polynomial expansion) and a small enough $\delta$ (around $1/20$) for the error, we guarantee in practice that the projection is indeed on the smallest $K-1$ eigenvectors.\\

\noindent
{\em Data preprocessing.}  
The MNIST dataset is composed of $n=60000$ images in the training set and 10000 images in the test set, where each sample is a black and white image of a handwritten digit of $28 \times 28$ pixels. The methodology is as follows: first, the dimension of the images is reduced with a PCA to something like 35 (or around 90, depending on the polynomial expansion degree we use in the following step). Fortunately, efficient incremental algorithms for PCA exist, where it is not required to fully diagonalize a covariance matrix, and the running time depends on the number of dimensions required as output. Second, a polynomial expansion of degree 2 or 3 is applied, hence making the dimension up to $10^4$. Third, the data is normalized so as to satisfy the SFA requirements of zero mean and unit variance. Overall, the preprocessing stage creates around $ n=10^5$ vectors $x(i), i\in [n]$ of size roughly $d = 10^4$ and the running time of the preprocessing is of the order of $\tilde{O}(nd)$, with $nd \approx 10^9$.  With a real quantum computer we would add a further step, which is to load the preprocessed data in the QRAM. This take only one pass over the data, and creates a data structure (i.e. a circuit) which is linear in the size of the data.  Hence, the overall preprocessing takes time of $\tilde{O}(nd)$.\\

\begin{figure}[h]
\centering
		\includegraphics[width=0.50\textwidth]{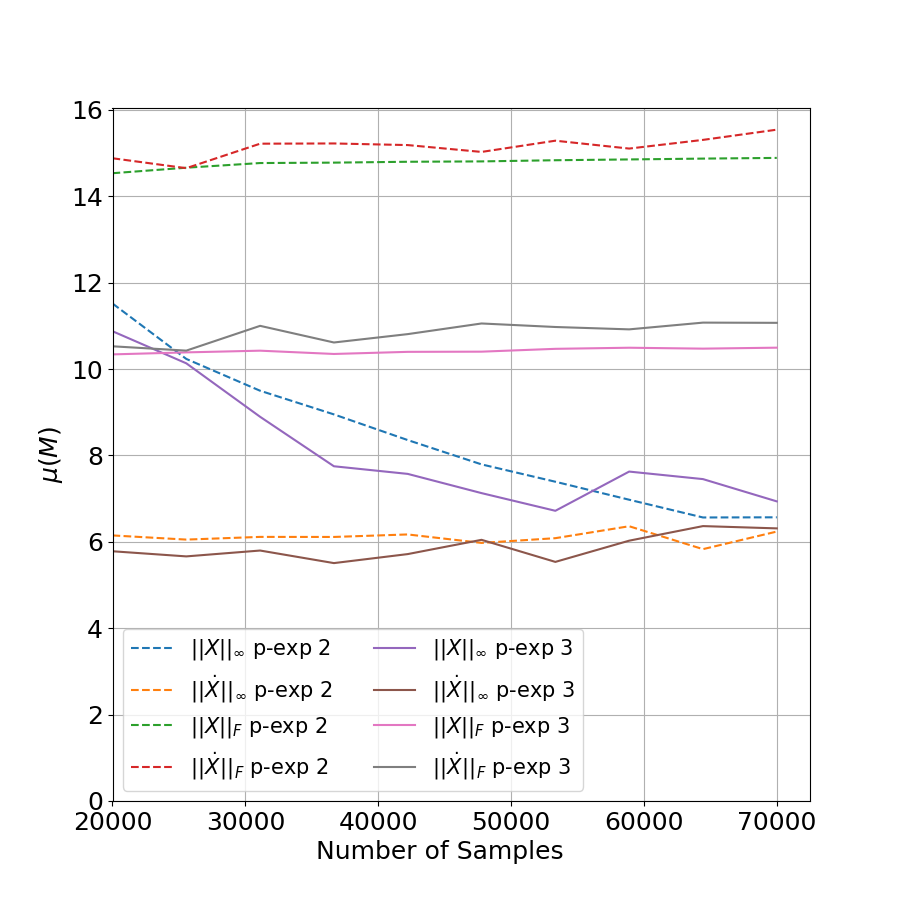}
			\caption{Sensitivity analysis of the parameter $\mu$ of the dataset while increasing $n$. The graph show the value of the Frobenius norm and the $\ell_\infty$-norm as two options for $\mu$. The two upper dotted lines represent the Frobenius norm of $X$ and $\dot{X}$ (respectively, lower and upper line) for polynomial expansion of degree 2. The thick horizontal lines above 10 shows the trend of the Frobenius norm for the polynomial expansion of degree 3 of $X$ and $\dot{X}$ (respectively, lower and upper line). The two lines at the bottom represent respectively the $\ell_\infty$ norm for $\dot{X}$ for for the polynomial expansion of degree 2 (dashed) and 3 (thick). The central slanting lines are the $\ell_\infty$ norm of $X$, for polynomial expansion of degree 2 (dashed), 3 (thick). 
			For the MNIST dataset, we see that both the Frobenius norm and the $\ell_\infty$ norm are practically constant when we increase the number of data points in the training set.
			}
	\label{fig_sensitivity_frob}
 		\end{figure}

\noindent
{\em Training.}
The classical SFA procedure outputs a small number $(K-1)$ of ``slow'' eigenvectors of the derivative covariance matrix, where $K$ is the number of different classes, and here $K=10$. This is in fact the bottleneck for classical algorithms and this is why the dimension was kept below $d=10^4$ with polynomial expansion, which still requires intensive HPC calculations. Generically, the running time is between quadratic and cubic, and hence of the order $10^{13}$. Once these eigenvectors are found, each data point is projected onto this subspace to provide $n$ vectors of $(K-1)$ dimensions which are stored in memory. As the points are labelled, we can find the centroid of each cluster. Note that at the end, the quantum procedure does not output a classical description of the eigenvectors, neither does it compute all vectors $y(i)$, as the classical counterpart could do. Nevertheless, given a quantum state $\ket{x(i)}$ it can produce the quantum state $\ket{y(i)}$ with high probability and accuracy. \\

\begin{figure}[h]
		\includegraphics[width=0.50\textwidth]{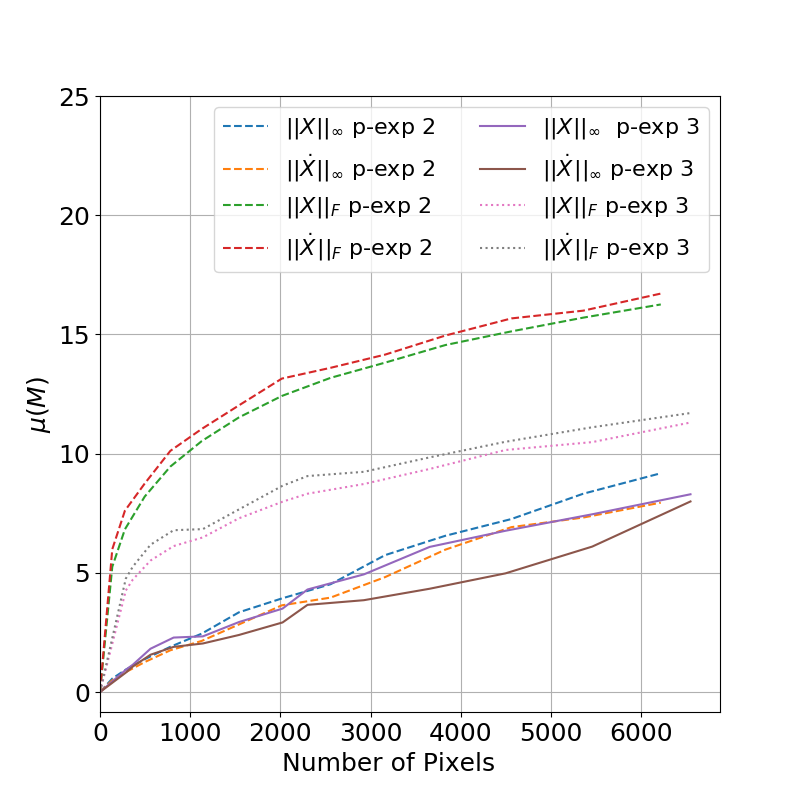}
	\caption{Sensitivity analysis of the parameter $\mu$ for the matrices $X$ and $\dot{X}$ while increasing $d$ - the number of pixels for an image - by changing the PCA dimension, which is performed before the polynomial expansion. The plot shows the value of the Frobenius norm and the $\ell_\infty$-norm (i.e. max  $\ell_1$ norm of the rows) as two options for $\mu$. The dashed lines are for the polynomial expansion of degree 2: the two upper lines are (from top to bottom) $\|\dot{X}\|_F$ and $\|X\|_F$. The two lower dashed lines are (from top to bottom) $\|X\|_\infty$ $\|\dot{X}\|_\infty$. 
	The two dotted central lines, track the Frobenius norm of the polynomial expansion of degree 3:  $\|\dot{X}\|_F$ and $\|X\|_F$ (from top to bottom). The bottom thick lines represent the $\ell_\infty$ norm for the polynomial expansion of degree 3 of $\|\dot{X}\|$ and $\|X\|$ (from bottom to top). In general, the $\ell_\infty$ norm is always smaller than the Frobenius norm. } 
	\label{fig_sensitivity_frob2}
\end{figure}

{\em Testing.}
For the testing stage, the classifier trained with the errors, is used to classify the 10000 images in the test set. Classically, the testing works as following: one projects the test data point $x(0)$ onto the slow feature space, to get a $(K-1)$-dimensional vector $y(0)$. Then, a classification algorithm is performed, for example kNN (i.e. one finds the $k$ closest neighbours of $y(0)$ and assigns the label that appears the majority of times). The complexity of this step is $O(Kd)$ for the projection of the test vector, plus the time of the classification in the slow feature space. The kNN algorithm, for example, is linear in the number of data points times the dimension of the points ($\tilde{O}(nd)$). In Nearest Centroid algorithm (a supervised classification algorithm where the label of a new point is assigned to the cluster with closest barycenter), if in the training stage we have found the centroids, then classification can be done in time $\tilde{O}(Kd)$. In our final algorithm we perform the training and testing together, i.e. using QSFA and QFD together. \\

\begin{figure}[h]
\includegraphics[width=0.5\textwidth]{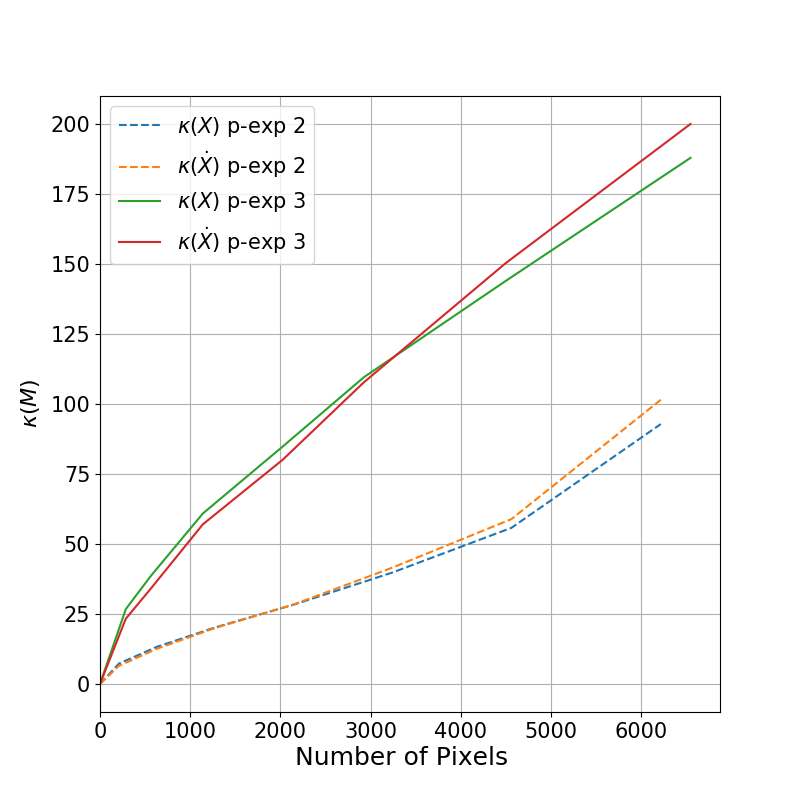}
	\caption{Sensitivity analysis of the condition number of $X$ and $\dot X$ while increasing $d$, the number of pixels, and discarding the 0 singular values. We plot the condition number of a polynomial expansion of degree 2: The upper dashed line is $\kappa(\dot{X})$, and the other dashed line is $\kappa(X)$. The upper thick lines are for a polynomial expansion of degree 3.  In both cases, the condition number of $\dot{X}$ dominates the condition number of $X$. For the case of a polynomial expansion of degree $3$, this happens after circa 3500 pixels.}
	\label{fig_sensitivity_cond1}
\end{figure}

\begin{figure}[h]
			\includegraphics[width=0.5\textwidth]{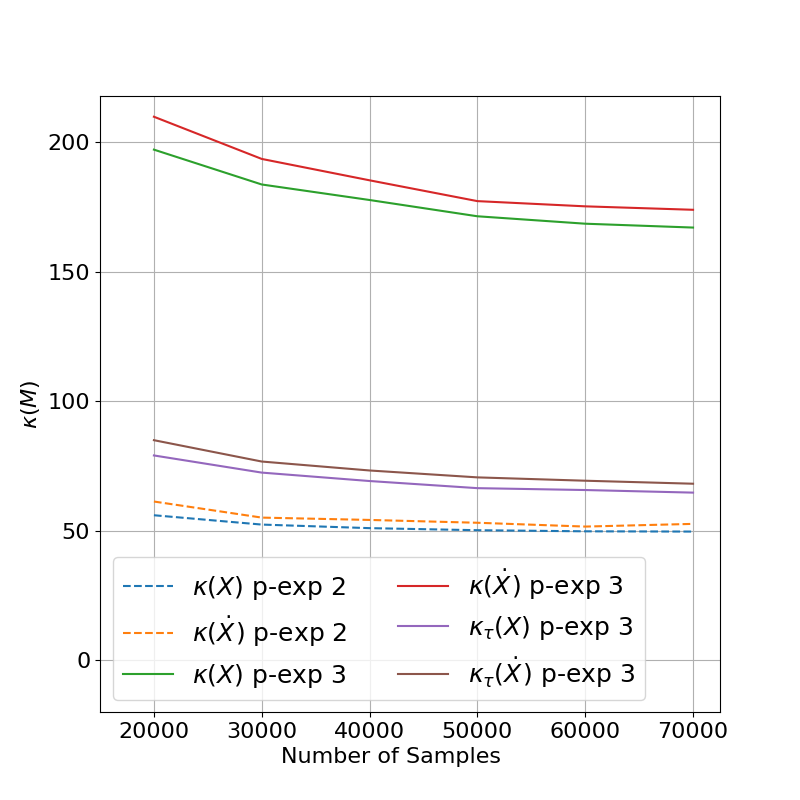}
		\caption{Sensitivity analysis of the condition numbers (after discarding the 0 singular values) while increasing the size of the training set (by adding the test set). The first two lines from the top represent the polynomial expansion of degree 3, with $\kappa(\dot{X})$ bigger than $\kappa(X)$. The two central lines are the condition numbers for the polynomial expansion of degree 3, while discarding $0.5\%$ of the smallest singular values (again $\kappa(\dot{X})$ is bigger than $\kappa(X)$). The two bottom lines represent the condition numbers for polynomial expansion of degree 2. In all three cases, the condition number of $\dot{X}$ is bigger than the condition number of $X$.}
		\label{fig_sensitivity_cond2}
	\end{figure}

\noindent
{\em Parameters of experiment.}
We now estimate a number of parameters appearing in the running time of the quantum classifier.\\

\noindent
{\em Number and dimension of data points.} For the MNIST we have that $nd$ is of the order of $10^9$ (including data points and derivative points). 

\noindent
{\em The parameter $\mu$ for the matrices $X$ and $\dot{X}$.}
We analyze the parameter $\mu(x)$, $\mu(\dot{X})$ as the number of data points in the training set and the dimension of the input vectors increases (PCA dimension + polynomial expansion). We know that $\mu$ is bounded by the Frobenius norm of the matrix. We also look at the case where $\mu$ is defined as the maximum $l_1$ norm of the rows of the matrices, plotted in Figure \ref{fig_sensitivity_frob}. Matrices are normalized to have spectral norm $1$. The good choice of $\mu$ is practically constant as we increase the number of points in the dataset. All the $l_1$ norms in the experiment were less than $11$. 
We also plot the Frobenius norm and the maximum $l_1$ norm as the dimension of the vectors in the dataset increases. While the Frobenius norm somewhat increases with the dimension, the maximum $l_1$ norm remains stable.
This could be expected since in the preprocessing a PCA is done, making the input matrices in fact quite low rank. Indeed, after the polynomial expansion the Frobenius norm does not increase much since we only add higher order terms, and note that all entries of the matrices are smaller than 1, since $\norm{X}_{max} \leq \norm{X}_2 \leq 1$. On the other hand, the scaling and normalization of $X$ helps keeping the $l_1$ norm even lower. 
It is important to state here that one gains a factor $10^3$ just by taking the correct quantum algorithm for performing linear algebra and not an off-the-shelf one. Such decisions will be crucial in reaching the real potential of quantum computing for machine learning applications.\\

\noindent
{\em Condition number for the matrix $X$.}
Figures \ref{fig_sensitivity_cond1} and \ref{fig_sensitivity_cond2} tells us that condition number is rather stable, in fact decreasing. As explained, we do not need to have the real condition number in the running time but a threshold under which we ignore the smaller eigenvalues. In fact, retaining just 99.5\% of the singular values does not considerably penalize the accuracy and achieves a behavior of growing much more slowly as we increase the dimension, with a value around $10^2$. \\

\noindent
{\em Error parameters.}
There are four error parameters, $\varepsilon$ for the matrix multiplication procedure, $\delta$ and $\theta$ for the projection procedure, and $\eta$ for the estimate of the norms in the classification. For $\varepsilon$, it appears only within a logarithm in the running time, so we can take it to be rather small. For the projection, we took $\delta \approx 1/20$ and from the simulations we have that $\theta \approx 0.3$ for polynomial expansion of degree 2 and $\theta \approx 0.05$ for polynomial expansion of degree 3. Last, it is enough to take $\eta=1/10$. Note also, that these parameters are pretty stable when increasing the dimension, as they only depend on whether we perform a polynomial expansion of 2 or 3.\\

\noindent
{\em Projection ratio.}
The ratio between the norm of the vectors in the whitened space over the projected vector in the slow feature space is well bounded. For an initial PCA dimension of 40 and polynomial expansion of degree $2$, this ratio is $10$ with variance $0.0022$, while for a polynomial expansion of degree $3$ and a PCA dimension of 30 is $20$ with $0.0007$ variance.\\

Table \ref{tableresults} provides the exact values of all parameters in the experiment.

\begin{table*}
	\caption{Accuracy of relevant experiments for various combination of classifiers and polynomial expansion. Here we have chosen $\epsilon/\kappa(X) = 10^{-7}, \delta =0.054, \eta =1/10$ and $10.000$ derivatives per class.}
	\label{tableresults}
	\begin{tabular}{|c|ccccccccccc|}
		
		$\text{QSFA}_2 $   & & & & & & & & & & &\\
		\hline
		d (PCA)      & $\norm{X}$ & $\norm{\dot{X}}$& $\norm{x(i)}_1$ & $\norm{\dot{x}(i)}_1$  & $\kappa(X)$ & $\kappa(\dot{X})$ &  $\kappa_t(X)$ & $\kappa_t(\dot{X})$ &  $\theta$ & \%$_t$ & \% \\
		\hline
		 860 (40)    & 22      &  102            & 0.9 & 1.4 & 41 & 22           &    32              &  15          & 0.38      & 96.4 & 96.4  \\
		3220 (80)   & 41   &    197          & 2.7 & 3.8  &  119       &  61          & 65                 &   30		         & 0.32      & 97.3 & 97.4 \\
		4185 (90)   & 46     & 215          & 3.1 & 4.4 & 143        &   74        & 72              &   35         & 0.31    &97.4  &97.4     \\
		\hline
		$\text{QSFA}_3$  & &  & & & & & & & & &\\
		\hline
		d (PCA)      & $\norm{X}$ & $\norm{\dot{X}}$& $\norm{x(i)}_1$ & $\norm{\dot{x}(i)}_1$  & $\kappa(X)$ & $\kappa(\dot{X})$ &  $\kappa_t(X)$ & $\kappa_t(\dot{X})$ &  $\theta$ & \%$_t$ & \% \\
		\hline
		5455 (30)   & 73     &    81          & 5.9 &4.3 & 276      & 278           &   149               &   149         & 0.06     & 98.2 &98.3    \\
		8435 (35)  & 96     &    102          & 7.8 & 5.7 &  369     & 389         &      146            &     156      & 0.05      & 97.5 & 98.4    \\
		9138 (36)   &  101    &  108     & 8.0 & 5.3 & 388       &  412       &   149               &    159       & 0.04      & 97.7 & 98.5      \\
	\end{tabular}
\end{table*}

\section{Construction of the QRAM} In this section we show how to construct the QRAM oracles needed in QSFA. QRAM has been used in quantum algorithmics literature as a generic way of retrieving classical data and build a corresponding quantum state. The name QRAM is meant to evoke the way classical RAM address the data in memory using a tree structure. A quantum query is defined as:
$$ \ket{i}\ket{0} \to \ket{i}\ket{b_i}  \mbox{ for } b_i \in  \mathbb{R}  \mbox{ and } i \in [N]$$

One, of course can write down the real $b_i$ with some precision $\delta$ using $\log 1/\delta$ bits.
We extend this data structure to allow us to efficiently create superpositions corresponding to the rows of the matrices, states with amplitudes equal to the norms of the rows of the matrices, and also states corresponding to the inputs that belong to a specific class $k$. We show what our QRAM data structure looks like for the input matrix $X$ (and similarly for th matrix $\dot{X}$) if the choice of $\mu$ is the Frobenius norm of the matrix. Each row of the matrix of the dataset is encoded as a tree, where the leaves correspond to the matrix elements, while the intermediate nodes store the square amplitudes that corresponds to their sub-tree.

We assume that the preprocessing (the optional PCA step, polynomial expansion, and removing the mean from the vectors and scaling the components to have unit variance) is performed classically, before storing the data in the QRAM. This takes $O(nd)$ time, which is upper bounded by $\tilde{O}(nd)$, the time needed to create this QRAM. Using the notation of the paper, we have $X \in \mathbb{R}^{n \times d}, X=\cup_{i=1}^K X_i$ and $X_k \in \mathbb{R}^{|T_k| \times d}$.  As we see in the figure, all rows of the matrix $X$ are saved as a tree with leaves the entries and root the corresponding norm $\norm{x(i)}$. We arrange the rows per class and we join all trees corresponding to rows of a class $k$ into a tree built on top of the individual trees, with leaves the norms of each row in the class (i.e. the roots of the previous trees) and root the norm of the class $\norm{X_k}$. On top of these $K$ trees, we built one last tree with leaves the norms of each class (i.e. the roots of the previous trees) and root the norm of the entire matrix $\norm{X}$. We also store the number of elements per class $T_k \in [K]$. This is said more coincisely in the following Corollary.
\begin{corollary}[QRAM for $X_k$]\label{quram_classes}
Let $X \in \mathbb{R}^{n \times d}$ and $X_k \in \mathbb{R}^{|T_k| \times d}$ for $k \in [K]$. There exists a data structure to store the rows of $X$ such that:
\begin{enumerate}
\item The size of the data structure is $O(n d\log^2(nd))$
\item The time to store a row $x(i)$ is $O(d \log^2(nd))$, and the time to store the whole matrix $X$ is thus $O(nd \log^2(nd))$	
\item A quantum algorithm that can ask superposition queries to the data structure can perform in time $polylog(nd)$ the following unitaries:
\begin{itemize}
\item $U: \ket{i}\ket{0} \to \ket{i}\ket{x(i)} $ for $i \in [n]$ 
\item $V: \ket{0} \to \sum_{i \in [n]} \norm{x(i)}\ket{i}$
\item $U_k: \ket{i}\ket{0} \to \ket{i}\ket{x(i)} $ for $i \in T_k$, for all $k \in [K]$ 
\item $V_k : \ket{0} \to \sum_{i \in T_k} \norm{x(i)}\ket{i}$, for all $k \in [K]$
\end{itemize}
\end{enumerate}
\end{corollary}

\begin{figure}[H]
  \begin{forest}for tree={math content, s sep=2mm, inner sep=0}
  [\norm{X} [\cdots    [\norm{X_1}   [\cdots 
  [\norm{x(1)} ] [\norm{x(2)} ]  ] [ \cdots [\norm{x(3)} ] 
  [\norm{x(|T_1|)} [\cdots [x_1(|T_1|) ] [x_2(|T_1|) ] ] 
  [\cdots [x_{d-1}(|T_1|) ] [x_d(|T_1|) ]  ]  ]]] 
  [\norm{X_2} ] ]     [\cdots  [\norm{X_{K-1}} [\cdots ] [\cdots ]] 
  [\norm{X_K}   [\cdots ] [\cdots ] ] ] ] 
  \end{forest}
  \caption{QRAM tree for matrices $X_k$ and for the Frobenius norms of the sub matrices of each class.}\label{img_qram}
  \end{figure}

%
%
%

The procedure to create and store the matrix $\dot{X}$ is exactly the same as the procedure needed for the $X$. Note that the classical matrix $\dot{X_k}$ is created in the following way: for each sample class $T_k$ in the training set, we are going to create $m=T_k log(T_k)$ new derivative vectors $\dot{x}(i) := x(i) - x(j) $ with $i,j \in T_k$ by sampling from the uniform distribution  $m$ pairs of vectors with the same label.

\section{Quantum Frobenius Distance classifier.}
We assume we can create a superposition of all vectors in the cluster as quantum states, and have access to their norms, which can be achieved either using our QRAM, or in case the quantum states are efficiently constructible by quantum circuits. We define $N_k= \norm{X_k}_F^2+ \norm{X(0)}_F^2 = \norm{X_k}_F^2 +|T_k|\norm{x(0)}^2$. We give the steps to build an efficient procedure to estimate distances as Algorithm \ref{QFE}, and later describe how to use it to build a classifier. For the analysis of the Algorithm \ref{QFE}, just note that the probability of measuring $\ket{1}$ in the final state is:
\begin{align}
\frac{1}{2N_k} \big ( |T_k|\norm{x(0)}^2 + \sum_{i \in T_k} \norm{x(i)}^2 -\nonumber \\ 2\sum_{i \in T_k} \braket{x(0), x(i)} \big) = F_k(x(0)).\end{align}
By Hoeffding bounds, to estimate $F_k(x(0))$ with error $\eta$ we would need $O(\frac{1}{\eta^2})$ samples.
For the running time, we assume all unitaries are efficient either because the quantum states can be prepared directly by some quantum procedure or given that the classical vectors are stored in the QRAM. Hence the algorithm runs in time $\tilde{O}(\frac{1}{\eta^2})$. We can of course use amplitude estimation and save a factor of $1/\eta$. Depending on the application or the hardware, one may prefer to keep the quantum part of the classifier as simple as possible or optimize the running time by performing amplitude estimation. Given this estimator, we can now define the QFD classifier. For a test point, the classifier simply runs the distance estimation procedure for each cluster of vectors with the same label. Then, the test point is assigned to the closest cluster. It is simple to see that the running time of this algorithm is $\tilde{O}(K/\eta^2)$ ( or $\tilde{O}(K/\eta)$ using amplification techniques).

\begin{algorithm}[H]
  \floatname{algorithm}{Algorithm}
  \caption{Quantum Frobenius Distance Estimator}
  \label{QFE}
  \begin{algorithmic}[1]
  \Require
  \Statex QRAM access to the matrix $X_k$ of cluster $k$ and to a test vector $x(0)$. Error parameter $\eta > 0$. 
  \Ensure
  \Statex $\overline{F_k( x(0))}$ s.t.  $| F_k(x(0)) - \overline{F_k( x(0))} | < \eta $. 
  \Statex
  \State s:=0
  \For {$r=O(1/\eta^2)$ }
  \State Create the state  
  $$ \frac{1}{\sqrt{N_k}}  \Big( \sqrt{|T_k|}\norm{x(0)}\ket{0} +\norm{X_k}_F \ket{1}\Big) \ket{0}\ket{0}$$
  \State Apply the unitary that maps: 
  $$\ket{0}\ket{0} \mapsto \ket{0} \frac{1}{\sqrt{|T_k|}} \sum_{i \in T_k} \ket{i}\; $$ and 
  $$  \ket{1}\ket{0} \mapsto \ket{1} \frac{1}{\norm{X_k}_F} \sum_{i \in T_k} \norm{x(i)} \ket{i}$$
  
  to the first two registers to get
  $$   \frac{1}{\sqrt{N_k}} \Big( \ket{0}  \sum_{i \in T_k}  \norm{x(0)} \ket{i} + \ket{1} \sum_{i \in T_k}  \norm{x(i)} \ket{i} \Big) \ket{0} $$
  \State Apply the unitary that maps
  $$ \ket{0} \ket{i} \ket{0} \mapsto   \ket{0} \ket{i} \ket{x(0)} \; \mbox{ and } \;   \ket{1} \ket{i} \ket{0} \mapsto   \ket{1} \ket{i} \ket{x(i)}$$
  
  
  \State Apply a Hadamard to the first register and measure it. If the outcome is $\ket{1}$ then s:=s+1
  
  
 
  \EndFor
  \State Output $\frac{s}{r}$.
  
  
  \end{algorithmic}
  \end{algorithm} 

\end{document}